\documentclass[10pt,showpacs,onecolumn]{revtex4-1}
\usepackage{color}
\usepackage{graphicx}
\usepackage{amsmath}
\usepackage{amssymb}
\usepackage{bm}
\usepackage{}
\usepackage[
bookmarks=true,
colorlinks,
linkcolor=blue,
urlcolor=blue,
citecolor=blue,
plainpages=false,
pdfpagelabels,
final,
breaklinks=true
]{hyperref} 
\usepackage{ulem}
\begin{document}

	\preprint{APS/123-QED}
	\title{High-purity valley-polarized currents induced by bichromatic optical fields in two-dimensional materials}
	\author{Wenqing Li$^{1}$}
\author{Xiaosong Zhu$^{1,3}$}\email{zhuxiaosong@hust.edu.cn}
\author{Liang Li$^{1}$}
\author{Wanzhu He$^{1}$}
\author{Jie Long$^{1}$}
\author{Pengfei Lan$^{1,3}$}\email{pengfeilan@hust.edu.cn}
\author{Peixiang Lu$^{1,2,3}$}\email{lupeixiang@hust.edu.cn}

\affiliation{%
		$^1$ Wuhan National Laboratory for Optoelectronics and School of Physics,
		Huazhong University of Science and Technology, Wuhan 430074,
		China\\
		$^2$ Hubei Key Laboratory of Optical Information and  Pattern Recognition, Wuhan Institute of Technology, Wuhan 430205, China\\
            $^3$ Hubei Optical Fundamental Research Center,  Wuhan 430074, China     }

	
	\begin{abstract}

    Producing currents predominantly from a single valley, namely valley-polarized currents, at optical-cycle timescales is an important aspect of the petahertz valleytronics, yet it remains less developed. 
    This work exhibits the feasibility of achieving this goal using bichromatic optical fields, which allow for the precise control of sub-cycle electron dynamics. 
    The combined effect of the helical and asymmetric waveforms of the optical fields leads to highly asymmetric excitation at different valleys and displacement of the excited electrons concurrently, thereby inducing valley-polarized currents with high valley purity, on the sub-optical-cycle timescale. 
    Inherently, the purity of the currents built up from the optical approach remains high even for materials with short decoherence time. 
    Moreover, the direction of the currents can be precisely controlled by adjusting the relative phase of the bichromatic components. 
    Our work offers a promising avenue for generating and modulating high-purity valley-polarized currents at the femtosecond timescale, facilitating the development of petahertz valleytronics.

	\end{abstract}
	
	\maketitle
	
	\section{INTRODUCTION}

        In the electronic band structure of a material, local minima in the conduction band (CB) or local maxima in the valence band (VB) are referred to as valleys.
        In addition to charge and spin, an electron also possesses a valley degree of freedom that indicates the specific valley it occupies. 
        The valley degree of freedom holds promise for energy-efficient information storage and presents intriguing prospects for going beyond classical information processing \cite{Yu2015, Schaibley2016, Vitale2018}.

        Utilizing the opposite Berry curvature associated with the different valleys, the manipulation of the valley polarization (the population imbalance at different valleys)
        has been achieved in diverse materials, \cite{Xiao2007, Mak2012, Xiao2012, DeGiovannini2016,  OliaeiMotlagh2018, JimenezGalan2020, JimenezGalan2021, Mrudul2021, Sharma2022, Sharma2023b,  Rana2023,  Tyulnev2024, Mitra2024} which is a key topic of valleytronics. 
        Since the valley index is a concept within the reciprocal space, the next pivotal challenge in valleytronic devices is how to transfer the valley information to the external environment \cite{Vitale2018}. 
        One promising solution is to generate and deliver the valley-polarized currents, where the carriers predominantly originate from a single valley. As a key component, it can be seamlessly integrated with other existing devices to form the foundation of future valleytronic devices \cite{Vitale2018, Lai2023}. 
        Over the past two decades, numerous efforts have been devoted to generating valley-polarized currents  \cite{Rycerz2007,  Zhang2014,  Gorbachev2014, Mak2014, Shimazaki2015, Jin2018, Jiang2013, Settnes2016, Gupta2019, Hsu2020}, including designing valley filters \cite{Rycerz2007} and valleytronic transistor \cite{Zhang2014}, utilizing the valley-Hall effect \cite{Gorbachev2014, Mak2014, Shimazaki2015} and employing advanced strain engineering techniques \cite{Jiang2013, Settnes2016, Gupta2019, Hsu2020}, etc. 
        These approaches rely on rigid architectures, such as a fixed pair of electrodes to supply a strong external bias or organized heterostructures \cite{Jin2018}.
        Recently, it is also proposed to generate valley-polarized currents with high valley purity utilizing strong terahertz (THz) pulses \cite{Sharma2023, Sharma2023a}, whose periods in experiments are typically on the picosecond scale. Despite all the above advances, the inherent characteristics of these configurations have precluded generating and modulating valley-polarized currents at petahertz (PHz) rates.

        To address this critical void, an optical approach to build-up a high-purity valley-polarized current is essentially required, where the applied fields oscillate at optical frequencies with femtosecond scale periods. 
        Recently, there is an abundance of studies investigating the manipulation of valley polarization \cite{OliaeiMotlagh2018, JimenezGalan2020, JimenezGalan2021, Mrudul2021, Sharma2022, Sharma2023b, Rana2023, Tyulnev2024, Mitra2024} and the injection of photocurrent  \cite{Schiffrin2012, Schultze2012, Higuchi2017, Heide2021, Langer2020, Neufeld2021, Rana2024, Weitz} using optical fields.
        For instance, the counter-rotating bicircular fields have been employed to induce valley polarization in various materials \cite{JimenezGalan2020, Mrudul2021, Tyulnev2024, Mitra2024}. Besides, it is proposed to generate photocurrents in a variety of solids with the co-rotating circular fields \cite{Neufeld2021}, where the direction of the currents can be controlled by the relative phase of the bichromatic components \cite{Neufeld2021, Rana2024, Weitz}. 
        To inject the photocurrents, one would induce an asymmetric electron population along the current direction throughout the Brillouin zone (BZ),  
        namely breaking the mirror symmetry of the population about the axis perpendicular to the current direction.
        Consequently, the corresponding injected photocurrents in hexagonal two-dimensional (2D) materials are often accompanied by a valley asymmetry \cite{customreference}. 
        However, without a specific design, the valley purity of the current is typically low, which hinders the effective transport of valley information and the development of corresponding applications. 
        Therefore, in the pursuit of valley-polarized currents, the term \textit{valley-polarized current} consistently refers to those with high valley purity \cite{Sharma2023, Sharma2023a, Abergel2009,  Isberg2013, Song2013, Zhang2014}.
        
        In this work, we demonstrate the generation and modulation of high-purity valley-polarized currents by applying the optical fields. It is realized by engineering the sub-cycle electron dynamics with co-rotating bicircular fields \cite{Milosevic2000, Fleischer2014, Kfir2014, Reich2016a, Han2018, Neufeld2021, Li2022, Zhu2022, Rana2022, Lerner2023, Rana2024, Weitz} that consist of a circularly-polarized fundamental field and its co-rotating second harmonic. 
        The combined effect of the helical and asymmetric waveforms of the fields concurrently leads to significant asymmetric excitation at different valleys and displacement of the excited electrons, giving rise to valley-polarized currents with high purity on the sub-optical-cycle timescale.
        Its feasibility is demonstrated by numerical simulations for the two typical categories of valley-active materials, gapped graphene and transition metal dichalcogenides (TMDs), based on a tight-binding model and the time-dependent density functional theory (TDDFT), respectively.
        Numerical results show that the purity of the optically built up valley-polarized currents can be only slightly degenerated by less than one-twelfth even when the decoherence time is reduced to 5 femtoseconds. 
        Besides, it is also shown that the direction of the currents can be precisely controlled by adjusting the relative phase of the bichromatic components. 
        Our work presents a promising approach for generating and modulating high-purity valley-polarized currents at the femtosecond timescale, facilitating the development of  PHz valleytronic devices.

        \section{RESULTS AND DISCUSSION}

        \subsection{Generation of valley-polarized currents via tailored optical fields}\label{schematical}

\begin{figure}[ht]
		\centerline{
		\includegraphics[width=14cm]{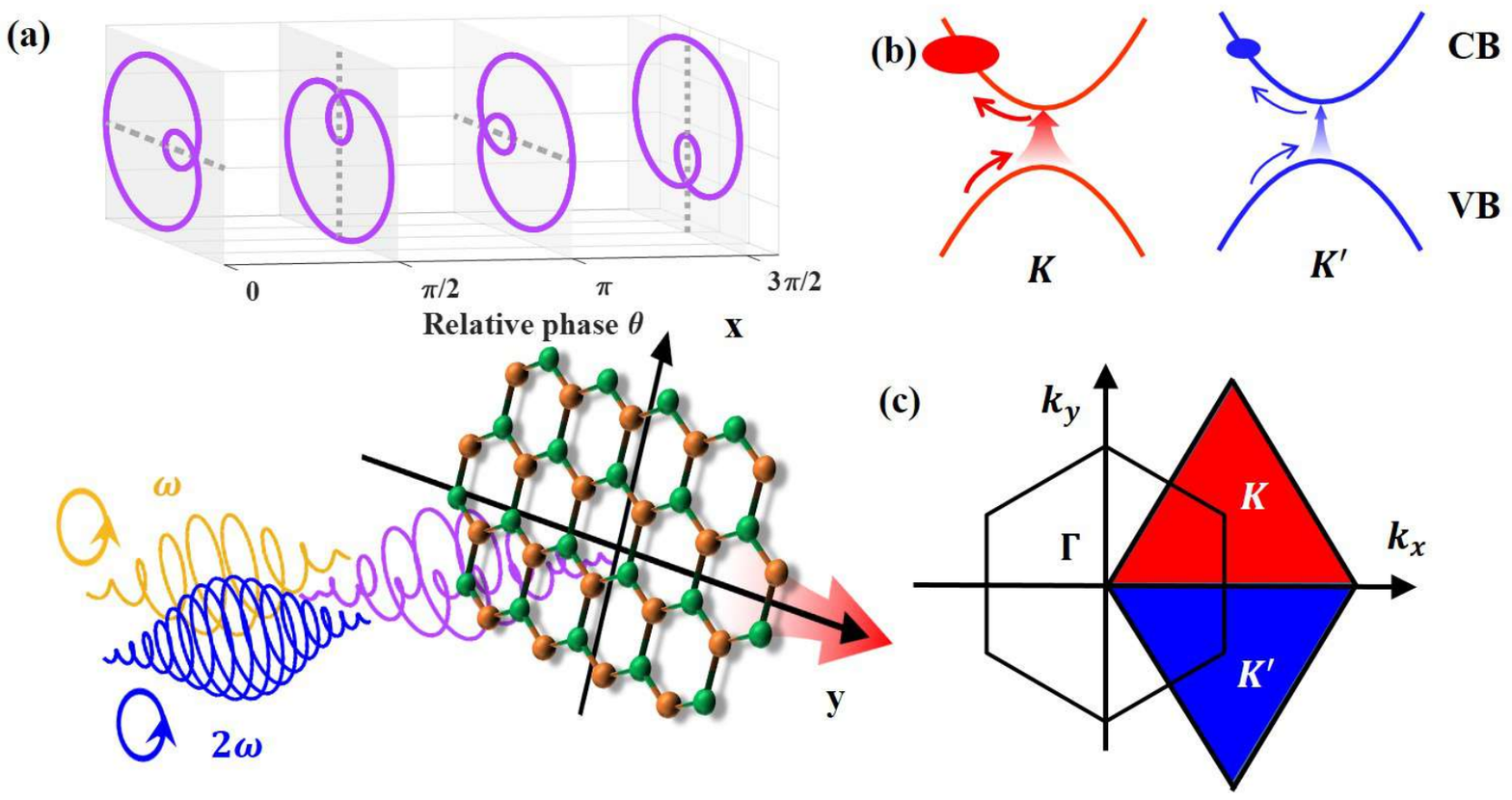}}
		\caption{\label{bandstructure}
			(a) The main panel schematically illustrates the generation and control of valley-polarized currents in 2D materials driven by the co-rotating bicircular field.  
            The upper panel shows the Lissajous figures of the co-rotating bicircular fields with different relative phases.
            (b) The schematic illustration of the excited electrons in the \textcolor{black}{ VB and CB} around two valleys. 
            (c) The first Brillouin zone (BZ) is separated into two regions around two valleys, as indicated by the red and blue regions.}
	\end{figure}

        Here, we outline the basic concept and offer guidelines for designing the tailored optical driving fields aimed at generating valley-polarized currents. 
     To obtain valley-polarized currents, two conditions are required: electron excitation predominantly takes place in one valley over the other, and the excited electrons are asymmetrically distributed around the valley. 
     Thus, the driving field must concurrently satisfy two crucial criteria. It should induce different electron dynamics in respective valleys and it must be capable of provoking the displacement of excited electrons.
     In this context, a co-rotating bicircular field, composed of a circularly polarized fundamental-frequency field and a frequency-doubling field with the same helicity, will fulfill the aforementioned requirements. 
     Its helical waveform drives distinct electron dynamics around different valleys with opposite Berry curvature. 
     Simultaneously, its asymmetric waveform displaces the excited population \cite{Neufeld2021} inducing the valley-polarized currents. 
     Considering the co-rotating driving laser is polarized in the $x-y$ plane and traveling in the $z$-direction as illustrated in the main panel in Fig.~\ref{bandstructure}(a), its electric field can be described by:
    \begin{equation}\label{co}
		\mathbf{F}_{\rm{co}}(t) = \frac{1}{\sqrt{2}} \left [ \begin{array}{l}
			F_{1}\cos (\omega t)+F_{2}\cos (2 \omega t + \theta) \\
			F_{1}\sin (\omega t)+F_{2}\sin (2 \omega t+ \theta)
		\end{array}\right ].
    \end{equation}
    $\omega$ is the frequency of the fundamental component.
    $F_{1}$ and $F_{2}$ are the amplitudes of each circularly-polarized component, corresponding to intensity $I_1$ and $I_2$, respectively.
    $\theta$ is the relative phase between the bichromatic components.
    Recent developments in modern light generation technology have enabled the customization of nontrivial optical waveforms by synthesizing lights with different colors \cite{ Fleischer2014, Kfir2014,  Han2018, Ge2019}, and thus such a field is readily available in experiments.

    For simplicity, we consider $F_{1} = F_{2} = F_{0}$.  
    Equation~(\ref{co}) can be factorized and rewritten as $\mathbf{F}_{\rm{co}}(t) = \sqrt{2} F_{0}\cos(\frac{\omega t + \theta}{2}) [\cos (\frac{3\omega t + \theta}{2}),\ \sin (\frac{3\omega t + \theta}{2})]^\top$, 
    which indicates that the $\omega-2\omega$ co-rotating driving field can be viewed as a circularly-polarized field modulated by $\cos(\frac{\omega t + \theta}{2})$ \cite{Reich2016, Li2022}. Then, one can obtain 
    \begin{equation}\label{F2}
		{F}^2_{\rm{co}}(t) = 2F_{0}^2\cos^2(\frac{\omega t + \theta}{2}) = F_{0}^2(1 + \cos(\omega t + \theta)).
    \end{equation}
    Namely, the magnitude of the field ${F}^2_{\rm{co}}(t)$ exhibits a cos-type modulation around 1, which indicates that the waveform of co-rotating bicircular fields (including not only the electric field but also the vector potential 
    $ \mathbf{A}(t) = -\int_{ \infty}^t\mathbf{F}(t')dt' $) is naturally asymmetric.
    The electric field reaches the maximum value at $\omega t = -\theta + 2\pi N$ ($N$ is an integer).
    Therefore, the symmetric axis of the corresponding Lissajous figure, which aligns with the maximum and minimum values of
    ${F}^2_{\rm{co}}$, as indicated by the dashed grey lines in the upper
    panel in Fig.~\ref{bandstructure}(a), is oriented along $\mathbf{{e}}_{\phi} = \cos(-\theta) \mathbf{{e}}_x + \sin(-\theta) \mathbf{{e}}_y$. Here, $\mathbf{{e}}_{x}$ ($\mathbf{{e}}_{y}$) represents the unit vector along the $x$ ($y$)-direction.
    $\mathbf{{e}}_{\phi}$ represents another unit vector, which makes an angle $\phi$ with $\mathbf{{e}}_x$. Here, $\phi = -\theta$.
    Therefore, the orientation of the waveform can be controlled by $\theta$, as illustrated in the upper panel in Fig.~\ref{bandstructure}(a).

    As shown in Fig.~\ref{bandstructure}(b), once the co-rotating bicircular field illuminates the 2D hexagonal materials lying in the $x-y$ plane, the electrons are pre-accelerated in the VB before being excited to the CB in the vicinity of the valleys  \cite{Li2019, Sun2021}. 
    The excitation predominately occurs when the electric field reaches its maximum of each optical cycle (e.g. see Fig.~\ref{graphene}(b)). After the excitation, the electrons continue to evolve in the CB with the driving field, occupying different regions in the BZ.
    The electron wavepacket evolves in the reciprocal space according to the Bloch acceleration theorem ($\mathbf{k}(t)=\mathbf{k}_{\rm{ex}}+\mathbf{A}(t)$, $\mathbf{k}_{\rm{ex}}$ is the electron wavenumber of the excited electron).
    Therefore, the vector potential $\mathbf{A}(t)$ is the quantity responsible for the asymmetric electron population symmetry in each valley (see Fig.~\ref{bandstructure}(b)), and
    the excited population is displaced along the negative direction of the maximum of $\mathbf{A}(t)$, which is similar to the current generation using a few-cycle laser pulse that arises from symmetry breaking \cite{Higuchi2017, Heide2021, Langer2020}.
    Consequently, the nonlinear residual photocurrents along the negative direction of the maximum of $\mathbf{A}(t)$ are obtained \cite{Neufeld2021, Rana2024} (for a particular example, see Fig.~\ref{graphene}). 
    Meanwhile, due to the helical waveform of the field synthesized by the two co-rotating components, the electrons around one valley can be dominantly excited, as indicated by the red and blue arrows. This can be understood by the fact that the electrons are driven by the optical field around respective valleys following the same reciprocal trajectories, including the same cycling direction. For electrons around the $\mathbf{K}$ and $\mathbf{K'}$ points, 
    they accumulate dynamical phases similarly, but geometric phases and transition dipole phases oppositely, which lead to growing or canceling electron populations for different valleys in CB \cite{Cui2022}. 
    As a result, the residual photocurrent dominantly originates from one valley only, thereby resulting in the valley-polarized currents. This can also be understood with the broken time-reversal symmetry of the co-rotating fields. 
    The process occurs within each optical cycle, which is a characteristic common to the manipulation of electron dynamics by light-wave \cite{Higuchi2017, Heide2021, Neufeld2021}, allowing great potential for generating and modulating valley-polarized currents in the sub-optical-cycle timescale. 

    According to the principle of our scheme, it primarily relies on the helical and asymmetric waveform of the optical field to initialize the valley polarization and induce the currents.  More fundamentally, it benefits from the breaking of symmetry \cite{Neufeld2021, Neufeld2019}, which are independent of the specific material properties. Thus, this scheme will apply to a broad range of valley-active materials. In the following sections, we will demonstrate its feasibility through numerical simulations using the two most widely employed valley-active materials, gapped graphene and TMDs, respectively.\\

    \subsection{Valley-polarized currents in gapped graphene}\label{gappedgraphene}

    Due to its unique electronic properties and high mobility, graphene has emerged as a promising material for valleytronic nanosystems \cite{Mrudul2021, Rycerz2007, Gorbachev2014, Jiang2013, Settnes2016, Gupta2019, Hsu2020, Shimazaki2015}. 
    In this subsection, we demonstrate the generation of valley-polarized current in the gapped graphene with $0.4\ \mathrm{eV}$ gap, which can be achieved by growing on an incommensurate substrate \cite{Nevius_2015}. 
    A recent work demonstrates that the tight-binding approach can provide an excellent description for the laser-induced electron dynamics in graphene. \cite{Li2021}. 
    Therefore, the corresponding electronic dynamics in gapped graphene is described by solving the density matrix equations based on a tight-binding model \cite{Vampa2014, Cui2022}. 
    Technical details are provided in Supporting Information.

    Firstly, to verify the aforementioned idea of generating valley-polarized currents, we consider an example set of laser parameters with the fundamental wavelength of 5000 nm and the total intensity $I = I_1 + I_2  =  1 \times  10^{9}\ {\rm W/cm^2}$, which is below the damage threshold of graphene \cite{Roberts2011}. The corresponding photon energy is below the band gap. The relative phase $\theta$ of the bichromatic components is 0, and the intensity ratio $I_2/I_1$ is 1.  
    The driving laser pulse is characterized by a sine-squared envelope with a full width with six optical cycles of the fundamental field. 
    A $35\ \mathrm{fs}$ decoherence time is used in this simulation.

    \begin{figure}
		\centering{
		\includegraphics[width=12cm]{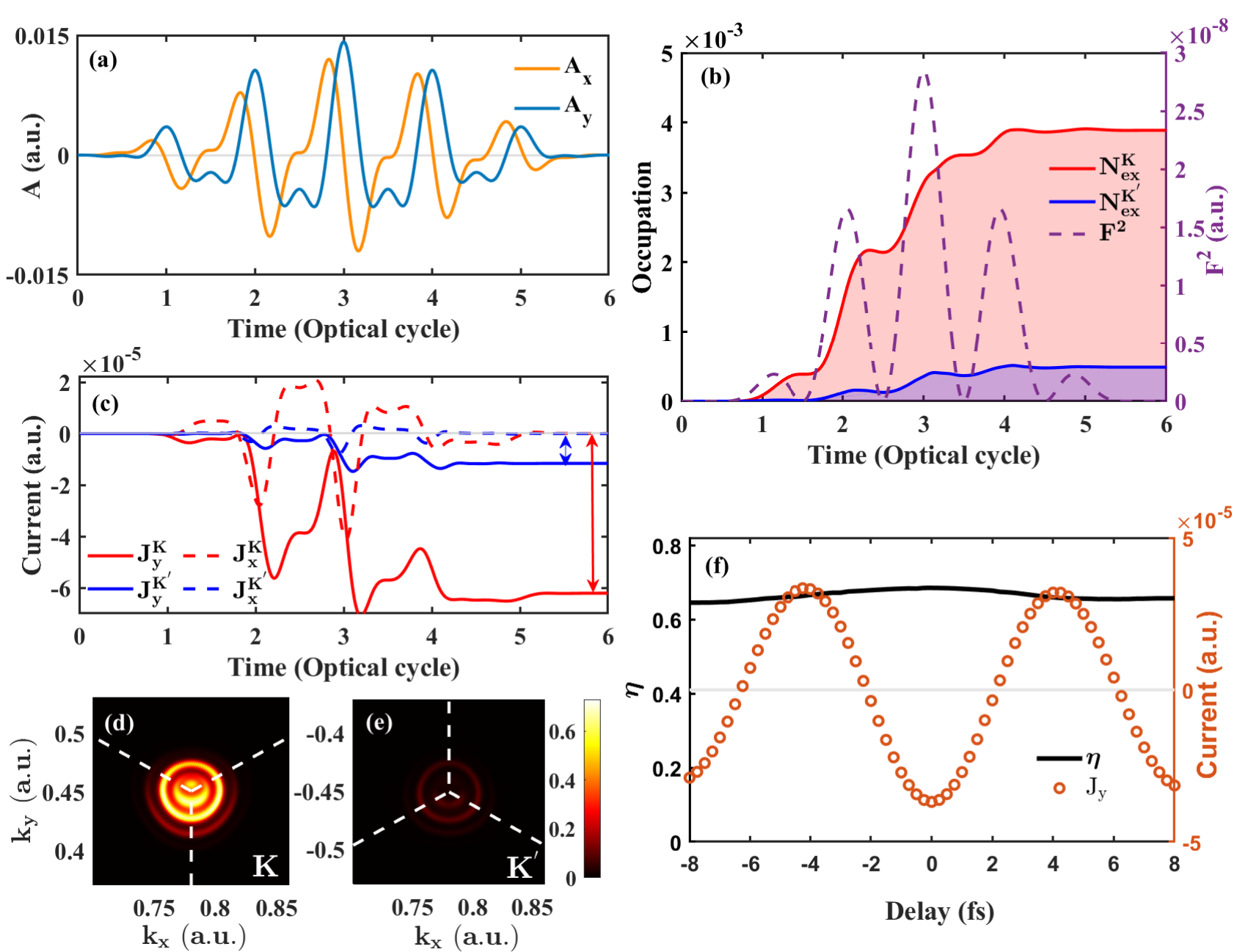}}
		\caption{\label{graphene}
			Valley-polarized currents in gapped graphene. 
            (a) The vector potential of the driving field. (b) The time evolution of electron occupation around two valleys in the CB and the square of the time-dependent electric field of the driving laser $F^2(t)$. (c) The $x$ and $y$ components of the currents from different valleys. (d)(e) The residual momentum-resolved electron population around $\mathbf{K}$ and $\mathbf{K^{'}}$ in CB, respectively. 
            (f) The dependence of the calculated ${J_y}$ and corresponding $\eta$ on the delay of the double-frequency component of the driving laser.}
	\end{figure}

                  \begin{figure*}
		\centerline{
		\includegraphics[width=14cm]{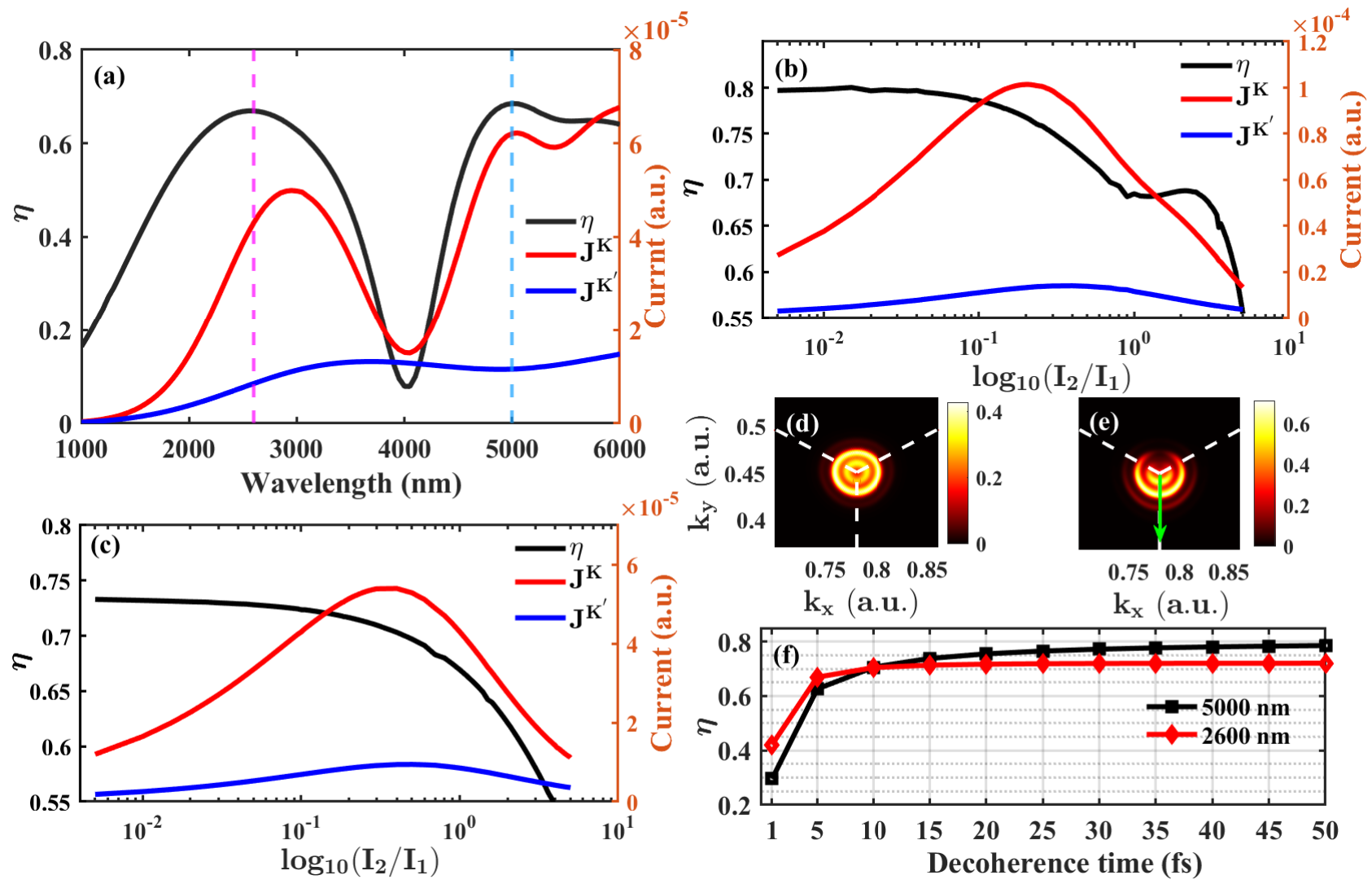}}
		\caption{\label{lambda_deco}
			(a) Fundamental wavelength dependence of the valley-polarized current generation. The optical parameters are the same as those for Fig.~\ref{graphene}, except for the fundamental wavelength.
                (b)(c) Intensity ratio dependence of the valley-polarized current generation for the fundamental wavelength of 5000 nm and 2600 nm, respectively.
                (d)(e) The residual momentum-resolved electron population around $\mathbf{K}$ point in CB with intensity ratio $I_2/I_{1} = 0$ and $0.15$, respectively. The corresponding fundamental wavelength is 5000 nm. The green arrows indicate the direction and magnitude of the valley-polarized currents.
                (f) Robustness of our scheme against decoherence considering the fundamental wavelength 5000 nm (black squares) and 2600 nm (red diamonds). The intensity ratio is $I_2/I_{1} = 0.15$.
                } 
	\end{figure*}

    Figure~\ref{graphene}(a) illustrates the $x$ and $y$ components of the time-dependent vector potential of the driving field. 
         Due to the helical waveform of the field, 
         the electrons around the $\mathbf{K}$ point are predominantly excited, as shown in Fig.~\ref{graphene}(b). 
         Thus, the electron population around $\mathbf{K}$ point (Fig.~\ref{graphene}(d)) is significantly larger than that of $\mathbf{K^{'}}$ point (Fig.~\ref{graphene}(e)). 
         Furthermore, the electron population in CB is displaced along the negative $k_y$ axis by the asymmetric $A_y$.
         Consequently, the intraband current along the negative $y$ axis is generated,
         as shown in Fig.~\ref{graphene}(c), consistent with the recent results \cite{Neufeld2021, Rana2024}.  
         Here, the intraband  current is calculated as:
        \begin{equation}\label{current}
            \begin{aligned}
           \mathbf{J}=  & \mathbf{J}^{K} + \mathbf{J}^{K^{'}} =  \\ & 
        \int_{\rm{BZ1}}2^2N_c(\mathbf{k})\frac{\partial E_c(\mathbf{k})}{\partial\mathbf{k}}d\mathbf{k} + \int_{\rm{BZ2}}2^2N_c(\mathbf{k})\frac{\partial E_c(\mathbf{k})}{\partial\mathbf{k}}d\mathbf{k},
       \end{aligned}
        \end{equation}

       where the $\rm{BZ1}$ ($\rm{BZ2}$) represents the region around $\mathbf{K}$ ($\mathbf{K^{'}}$), as shown in  Fig.~\ref{bandstructure}(c), $E_c$ denotes energy of CB, and $N_c(\mathbf{k})$ is the electron population on  CB. 

       Combined with the fact that there are more electrons excited in the vicinity of the $\mathbf{K}$ point (see Figs.~\ref{graphene}(d)(e)), the total current $\mathbf{J}$ dominantly originates from $\mathbf{K}$ point (see the red and blue double arrows in Fig.~\ref{graphene}(c)). Namely, a highly valley-polarized current is obtained.
       Its valley purity, evaluated by $\eta=(|\mathbf{J}^{\mathbf{K}}|-|\mathbf{J}^{\mathbf{K^{'}}}|)/(|\mathbf{J}^{\mathbf{K}}|+|\mathbf{J}^{\mathbf{K^{'}}}|)$, is $\eta = 68 \%$. 
       For the pristine graphene, recent works \cite{Rana2024, Weitz} have demonstrated that the co-rotating fields lead to photocurrents with nearly zero valley purity. 
       Besides, as shown by the solid red and blue lines in Fig.~\ref{graphene}(c), the valley-polarized current merges in one optical cycle. 
       This indicates that this scheme also works for shorter few-cycle laser pulses.

       In Fig.~\ref{graphene}(f), we demonstrate the control of valley-polarized currents by delaying the double-frequency component, which allows for fine-tuning of the driving optical field and precise control of the sub-cycle electron dynamics \cite{Schiffrin2012, Guan2020}.
       Accordingly, the current along the $y$ direction exhibits an oscillation with a period of 8 fs,  signifying a pivotal advancement toward PHz valleytronics. 
       Besides, the calculated $\eta$ shows good robustness to the delay.

       Having validated the basic concept, we will next explore strategies for optimizing the valley-polarized currents by altering the parameters of the driving field.    
       In  Fig.~\ref{lambda_deco}(a), the fundamental wavelength dependence of the valley-polarized current generation is depicted. The optical parameters remain consistent with those in Fig.~\ref{graphene}, except for the fundamental wavelength. 
         Within the range of $1000-6000\  \mathrm{nm}$, the current from $\mathbf{K}$ point (the red line) is significantly larger than the current from $\mathbf{K^{'}}$ point (the blue line), 
         which confirms the generality of the scheme over the laser wavelengths.
         Consequently, the calculated $\eta$, depicted by the black line, reaches the maximum value at $2600\ \mathrm{nm}$ and $5000\ \mathrm{nm}$ (dashed pink and light blue line).  
         The calculated $\eta$ is $67 \%$ and $68 \%$, respectively.

         By varying the intensity ratio $I_2/I_1$ of bicircular fields, one can achieve significant control over the electron dynamics \cite{Milosevic2000, Zhu2022, Mrudul2021a}, making it possible to enhance the purity of the generated valley-polarized currents.
        The impact of intensity ratio $I_2/I_1$ on the valley-polarized current generation with the optimal fundamental wavelengths $5000\ \mathrm{nm}$ and $2600\ \mathrm{nm}$ is shown in Figs.~\ref{lambda_deco}(b)(c), respectively.
        In the limit of $I_2 = 0$, the vector potential $\textbf{A}(t)$ is symmetric, thus the residual electron population is symmetric around $\mathbf{K}$ point, resulting in the absence of residual current, as shown in Fig.~\ref{lambda_deco}(d).
        With the increase of intensity ratio, the symmetry breaking of $\textbf{A}(t)$ increases and finally decreases (for the cases $I_2/I_1 = 0$ and $I_2/I_1 = \infty $, $\textbf{A}(t)$ is symmetric).
        Thus, the symmetry breaking of the electron population around each valley in the CB increases and then decreases. 
        Besides, the residual electron population around each valley is almost unchanged for different intensity ratios (not shown).
        Thus, the currents from the two valleys also increase and then decrease with the increase of intensity ratio, as the red and blue lines show in Figs.~\ref{lambda_deco}(b)(c).
        Regarding the purity $\eta$, it is highest for small $I_2/I_1$ (the left regions in Figs.~\ref{lambda_deco}(b)(c)). However, the generated currents from the two valleys are both small in these cases, as the weak 2$\omega$ component causes only minimal symmetry breaking in the electron population around each valley.        
        Thus, this configuration proves inadequate for the effective valley-polarized current generation. 
        For the case of $I_2/I_1 = 0.15$, the purity of obtained valley-polarized current is $78 \%$ and $72 \%$ for $5000\ \mathrm{nm}$ and  $2600\ \mathrm{nm}$, respectively. 
        Meanwhile, the generated currents are around their maximum values. 
        This can be regarded as the optimal intensity ratio.
        The corresponding residual electron population around $\mathbf{K}$ is shown in Fig.~\ref{lambda_deco}(e), exhibiting a pronounced asymmetric distribution. 

        Next, we examine how the purity $\eta$ of the optically built up valley-polarized currents is sensitive to the decoherence time in Fig.~\ref{lambda_deco}(f). Here, $\eta$ is calculated with different decoherence times and the optimized laser parameters from Figs.~\ref{lambda_deco}(a)(c) ($I_2/I_{1} = 0.15$ and fundamental wavelengths with 5000 nm and 2600 nm).  
        For the case of 5000 nm (2600 nm),  one can see that $\eta$ decreases from $78 \%$ ($72 \%$) to $63 \%$ ($67 \%$), 
        as the decoherence time reduces from $50\ \mathrm{fs}$ to $5\ \mathrm{fs}$.
        Even for the extremely short decoherence time of $1\ \mathrm{fs}$, this scheme still works, yielding the $\eta$ around $30 \%$ for 5000 nm ($42 \%$ for 2600 nm). 
        This feature arises from the fact that the valley-polarized currents are established based on the ultrafast electron dynamics. 
        Thus, this scheme allows the establishment of noticeable valley-polarized current for the materials with short decoherence time, like Weyl semimetals (a few femtoseconds, see \cite{Bharti2022, Medic2024, Huebener2017}) and Dirac semimetals (for instance, around $10\ \mathrm{fs}$ for a typical 3D Dirac semimetal $\mathrm{Cd_3As_2}$ \cite{Kovalev2020, Germanskiy2022}). It is worth noting that the established current also undergoes decoherence processes during the nonlocal transportation, which are not included in our simulations.
         This issue could be overcome through the design of micro-nano structures \cite{Gorbachev2014, Mak2014, Higuchi2017}. 

         \begin{figure}
		\centerline{
		\includegraphics[width=8.5cm]{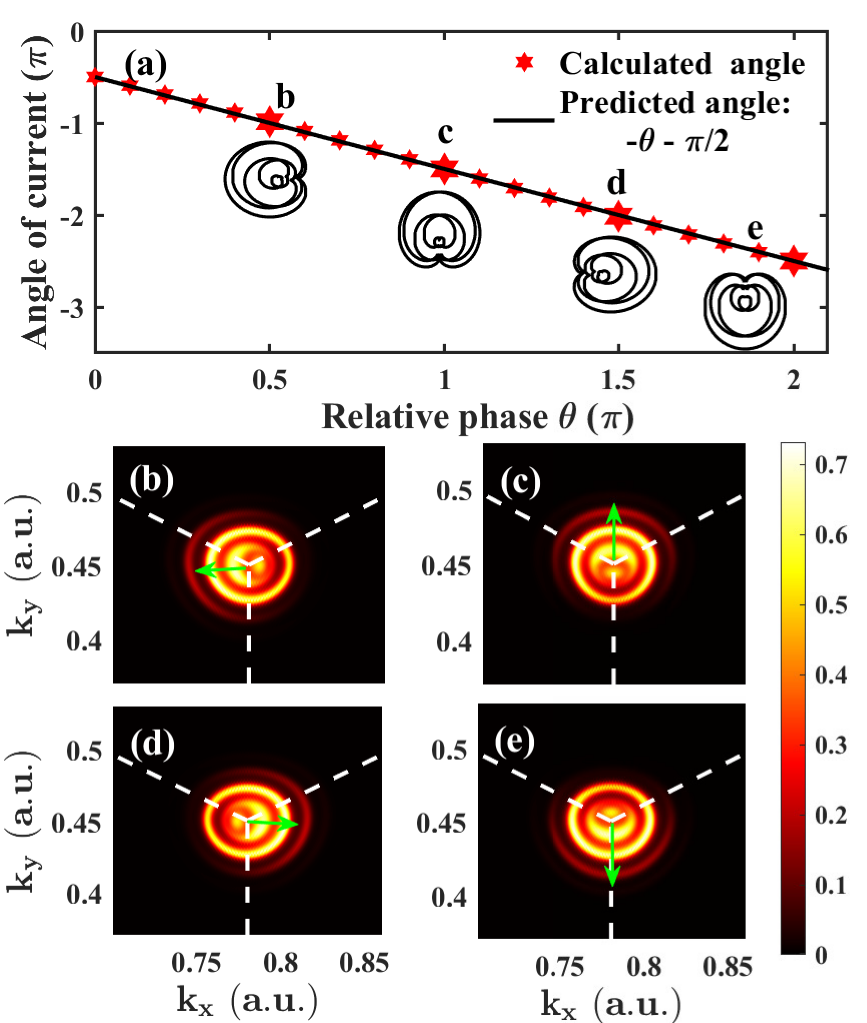}}
		\caption{\label{angle}
			(a) The red stars represent the direction of the calculated valley-polarized current with different relative phases $\theta$. The solid black line indicates the predicted direction of the valley-polarized current. The inserts show the waveforms of co-rotating bicircular fields ($-\mathbf{A}(t)$) for $\theta = \pi/2$,$\pi$, $3\pi/2$, and  $2\pi$ labeled by the big red stars b, c, d, and e, respectively. (b)-(e) The residual momentum-resolved electron population around $\mathbf{K}$ point in CB for the relative phase $\theta = \pi/2$, $\pi$, $3\pi/2$, and  $2\pi$, corresponding to the big red stars b-e in (a). The green arrows indicate the direction and magnitude of the valley-polarized currents.} 
	\end{figure}

                         \begin{figure*}
		\centerline{
		\includegraphics[width=14cm]{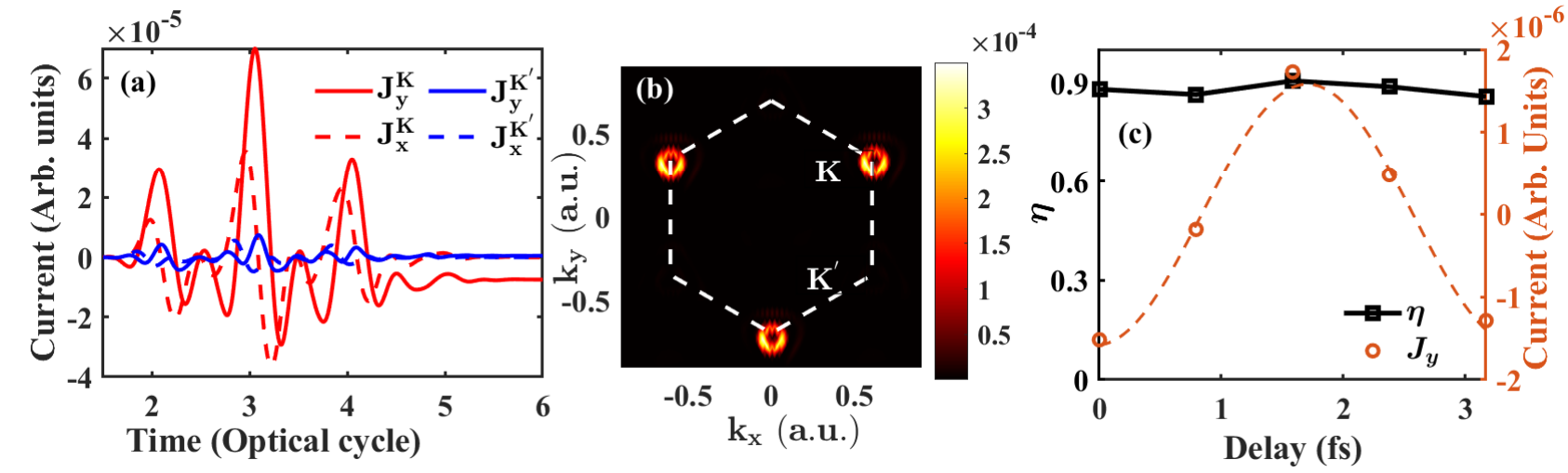}}
		\caption{\label{MoS2}
		 Valley-polarized currents in $\rm{MoS_2}$. (a) The $x$ and $y$ components of the currents from different valleys. (b) The residual momentum-resolved electron population in CB. (c) The dependence of the calculated ${J_y}$ and corresponding $\eta$ on the delay of the double-frequency component of the driving laser. The dashed orange line is drawn to guide the eye. } 
	\end{figure*}

    Another advantage of the all-optical approach based on the tailored bichromatic optical field is the extremely high degree of freedom, allowing versatile control of the valley-polarized currents.
       Finally, we demonstrate the precise modulation of the direction of valley-polarized current by adjusting the relative phase $\theta$.
       According to the discussion based on Eq.~(\ref{F2}), the symmetric axis of the electric field of the co-rotating bicircular field is oriented at  $\mathbf{{e}}_{-\theta}$, along which the Lissajous figure is asymmetric.  
       Likewise, one will find for the vector potential $ \mathbf{A}(t) = -\int_{ \infty}^t\mathbf{F}(t')dt' $ that the symmetric axis is oriented at $\mathbf{{e}}_{-\theta + \pi/2}$, differing from that of the electric field by $\pi/2$. 
       Consequently,
       it is expected that the residual electron population is displaced along $\mathbf{{e}}_{-\theta - \pi/2}$, which induces the residual valley-polarized current with the angle of ${-\theta - \pi/2}$, as shown by the solid black line in  Fig.~\ref{angle}(a). 
       
       To verify the above discussion, we numerically simulate the valley-polarized current generation with different relative phases. The directions of the resultant currents are depicted by the red stars in Fig.~\ref{angle}(a). 
       The optical parameters are the same as Fig.~\ref{graphene}(a), except for the relative phase $\theta$. 
       One can see that the numerical results perfectly match the prediction, which is consistent with the symmetry analysis and numerical simulations in the recent works \cite{Neufeld2021, Neufeld2019}.
       The corresponding residual momentum-resolved electron population around $\mathbf{K}$ and the magnitude of the residual currents (the green arrows) for four representative cases with $\theta = \pi/2$, $\pi$, $3\pi/2$, and  $2\pi$ are shown in Figs.~\ref{angle}(b)-(e), respectively.
       One can see that the magnitude remains nearly unchanged with the variation of $\theta$, which allows ideal control over only the direction of valley-polarized currents without changing their magnitude.

       \subsection{Valley-polarized currents in M\lowercase{o}S$_2$}

       Having demonstrated the fundamental physics and feasibility of the scheme with gapped graphene using a tight-binding model, we now examine its applicability to TMDs. 
        The TMDs are the other promising class of material extensively explored for valleytronics. 
        Compared with graphene, 
        the TMDs have broader band gaps and stronger multielectron effects \cite{Liu2016, ChangLee2024}.
        Thus, the valley-polarized current generation in TMDs is studied using state-of-the-art TDDFT simulations. This method allows one to model the electron dynamics in solids without making strong assumptions and has been shown to provide good agreement between simulation results and experimental measurements \cite{Klemke2019, Schumacher2023}. 
        The TDDFT simulations are performed with the real-space grid-based code, Octopus \cite{Marques2003, Castro2006, Andrade2015, Huebener2018, TancogneDejean2020}.
        Technical details are provided in Supporting Information.

    We take $\rm{MoS_2}$ as a prototype. For the driving field, the fundamental wavelength is 1900 nm and the total intensity is $1 \times  10^{11}\ {\rm W/cm^2}$, which is below the damage threshold of $\rm{MoS_2}$ \cite{Liu2016}.  
        The other optical parameters are consistent with those in Fig.~\ref{graphene}(a). 
        The waveform of the driving field is similar to Fig.~\ref{graphene}(a), and $A_y$ is still asymmetric.
        The result is shown in Fig.~\ref{MoS2}(a). One can see that the magnitude of the current originating from the $\mathbf{K}$ point ($J_y^{\mathbf{K}}$) is much higher than that from the $\mathbf{K'}$ point ($J_y^{\mathbf{K'}}$). Thus, a significant valley-polarized current is obtained along the negative $y$ direction with the purity of $88 \%$. As previously discussed, the emergence of the nonvanishing current results from the displacement of the electron population due to the asymmetric driving field and the high purity results from the distinct excitation rates at different valleys (see Fig.~\ref{MoS2}(b)). 
        We also calculate the current in $\rm{MoS_2}$ by delaying the double-frequency component, as shown in Fig.~\ref{MoS2}(c). Similarly, the current along the $y$ direction oscillates with a period of 3.5 fs while $\eta$ remains robust to the variation of the delay. 
        These results suggest that this approach is model-independent, cross-validating its feasibility across a wide range of valley-active materials. \\

	\section{CONCLUSION}

        In conclusion, we theoretically demonstrate the production of high-purity valley-polarized currents at the optical frequency level. 
        The applied co-rotating bicircular driving fields with helical and asymmetric waveforms concurrently induce significant valley polarization and population displacement in 2D materials on the sub-optical-cycle timescale.
        This approach is primarily dependent on the symmetry breaking, induced by the waveform of the driving field, and thus should apply to a broad range of valley-active materials. The feasibility is further supported by numerical simulations on the two typical categories of valley-active materials: gapped graphene and TMDs. 
       Numerical results show the purity of the optically built-up currents only slightly decreases even when the decoherence time falls to 5 fs. 
       Note that, although multi-cycle optical pulses are employed in the numerical demonstration, the scheme also works for even shorter few-cycle pulses as the valley-polarized currents stem from the sub-cycle electron dynamics.
    In addition, the direction of the currents can be precisely controlled by adjusting the relative phase of the bichromatic fields, with the magnitude of the current nearly unchanged.
    Our work demonstrates a promising avenue for generating and modulating high-purity valley-polarized currents at the femtosecond timescale, thereby promoting the PHz valleytronics.

        \section*{ACKNOWLEDGMENTS}
	This work was supported by National Key Research and Development Program (Grant No. 2023YFA1406800 and No. 2022YFA1604403) and the National Natural Science Foundation of China (NSFC) (Grant Nos. 12174134, 12104172, 12021004 and  12374317). The computation is completed in the HPC Platform of Huazhong University of Science and Technology.

	\bibliography{liwenqing}
\end{document}